# Magnetic proximity and nonreciprocal current switching in a monolayer WTe$_2$ helical edge

Wenjin Zhao[1], Zaiyao Fei[1], Tiancheng Song[1], Han Kyou Choi[1], Tauno Palomaki[1], Bosong Sun[1], Paul Malinowski[1], Michael A. McGuire[3], Jiun-Haw Chu[1], Xiaodong Xu[1,2*], David H. Cobden[1*]

[1]Department of Physics, University of Washington, Seattle, WA, USA
[2]Department of Materials Science and Engineering, University of Washington, Seattle, WA, USA
[3]Materials Science and Technology Division, Oak Ridge National Laboratory, Oak Ridge, TN, USA

*Corresponding authors: cobden@uw.edu, xuxd@uw.edu

## Abstract:

The integration of diverse electronic phenomena, such as magnetism and nontrivial topology, into a single system is normally studied either by seeking materials that contain both ingredients, or by layered growth of contrasting materials[1-9]. The ability to simply stack very different two-dimensional (2D) van der Waals materials in intimate contact permits a different approach[10,11]. Here we use this approach to couple the helical edges states in a 2D topological insulator, monolayer WTe$_2$[12-16], to a 2D layered antiferromagnet, CrI$_3$[17]. We find that the edge conductance is sensitive to the magnetization state of the CrI$_3$, and the coupling can be understood in terms of an exchange field from the nearest and next-nearest CrI$_3$ layers that produces a gap in the helical edge. We also find that the nonlinear edge conductance depends on the magnetization of the nearest CrI$_3$ layer relative to the current direction. At low temperatures this produces an extraordinarily large nonreciprocal current that is switched by changing the antiferromagnetic state of the CrI$_3$.

## Main Text:

The introduction of magnetic order into topological band structure gives rise to new phenomena such as the quantum anomalous Hall effect[1,8,9] and nonreciprocal magnetoelectric effects[5-7,18]. In the case of a two-dimensional topological insulator (2D TI), topology guarantees the existence of helical edge modes in which the spin is locked to momentum, causing current at the edge to be fully spin-polarized (the quantum spin Hall effect)[19]. Combining 2D TIs with magnets should therefore directly yield magnetoelectric coupling[20,21]. For example, a magnetic proximity effect may modify the spin polarization and hence the current, or the flow of current in the edge may produce a torque on the magnetization[22,23]. Since backscattering in the edge modes is suppressed by time reversal symmetry, the edge conduction should be affected by magnetic order, which will mix the two opposite-spin branches and so modify backscattering. The expected gapping of the helical edge modes by proximity with a ferromagnet is an important way to control them which, combined with induced superconductivity[24,25], plays a role in schemes to produce Majorana modes[26].

Stacking van der Waals materials offers a simple, flexible and low-disorder approach to combining magnets with other materials[10,11]. In this work we measure transport through a 2D TI, monolayer (1L) WTe$_2$[13-16], stacked under the layered magnetic insulator CrI$_3$[17,27-30]. We find that the magnetism of the CrI$_3$ suppresses the edge conduction in the WTe$_2$, in a manner consistent with the opening of a gap by an exchange field. The linear edge conductance is sensitive to the magnetization state of the CrI$_3$, and changes suddenly when the magnetization of the nearest or next-nearest CrI$_3$ layer flips. In addition, the nonlinear current-voltage characteristic has an even component that changes sign when the magnetization of the adjacent CrI$_3$ layer reverses. This is related to the "unidirectional magnetoresistance" seen in magnetically doped 3D TI structures where it was explained by spin-flip scattering of electrons by magnons[5,6]. However, in the helical edge at low temperatures the effect exists at zero external field and can be extremely large, creating a difference in (nonreciprocal) current of order 100% between the two opposite antiferromagnetic ground states of the CrI$_3$.

The structure and magnetic configuration of CrI$_3$ at zero magnetic field is indicated in Fig. 1a. Each layer is internally ferromagnetic, with the moments aligned out of the plane below a critical temperature of $T_C \approx 45$ K. In thin exfoliated CrI$_3$ flakes, adjacent layers are antiferromagnetically coupled and thus have opposite magnetization. The structure of 1L WTe$_2$ is sketched in Fig. 1b. Although semi-metallic at room temperature or when doped, below ~100 K and at low gate voltages the 2D bulk shows insulating behavior while the edge remains conducting. Both theory[12] and experiments[13-16] indicate that the edge states are helical (signified by the green and pink bordering lines representing the two spins channels), i.e., this is a quantum spin Hall system. For example, the edge conductance is strongly suppressed by an in-plane magnetic field. Each of our devices contains a monolayer flake of WTe$_2$ that is either partly or completely covered by a few-layer flake of CrI$_3$, as sketched in Fig. 1c. The thicknesses of monolayer WTe$_2$ and few-layer CrI$_3$ were determined from optical contrast[13,17] (Supplementary SI-1). The WTe$_2$ overlies prepatterned platinum contacts, all encapsulated between two hexagonal boron nitride (hBN) dielectric layers, with a graphite bottom gate to which a gate voltage $V_g$ is applied (see Methods and Supplementary SI-1).

Figure 1d shows the $V_g$ dependence of the linear-response conductance $G$ measured between two adjacent contacts in device C1, which has trilayer CrI$_3$ covering most of the WTe$_2$ (see the inset optical image). $G$ exhibits a minimum near $V_g = 0$. The value at the minimum, plotted vs $T$ in Fig. 1e (black points), decreases monotonically on cooling. Above about 50 K the behavior is similar to that of a typical bare 1L WTe$_2$ device M1 (red points), in which the bulk conductivity steadily decreases on cooling. However, at lower temperatures in the bare device the conductance levels off due to temperature-independent edge conduction, whereas in C1 it continues to drop to the lowest temperature of 5.6 K. Similar behavior was seen in four devices whenever CrI$_3$ covered at least part of the edge in the current path (Supplementary SI-2). The blue dashed line is an Arrhenius fit of the form $G \propto \exp(-E_a/k_B T)$ to the data in this regime, where $k_B$ is Boltzmann's constant, yielding an activation energy $E_a = 2.5 \pm 0.3$ meV. This activated suppression of the edge conduction is similar to the behavior seen in an applied magnetic field, where the activation energy is found to be approximately proportional to the field[13]. However, a theory of this suppression that takes into account the combined effects of wavefunctions at the edge, disorder, magnetism, and electron-electron interactions is not yet available. We therefore focus here on the effects of changing the magnetization state of the CrI$_3$ which we do using a perpendicular applied field $B$.

Figure 2a shows the linear conductance of device C1 as $B$ is swept upwards (orange) and downwards (green) at a series of temperatures, measured at a gate voltage ($V_g$ = -0.5 V) where edge conduction dominates. To within the noise the upwards traces are the same as the downwards traces reflected in $B = 0$, as should be the case by time-reversal considerations. At temperatures above $T_c$~45 K, $G$ decreases smoothly with increasing $B$ as does the edge conduction in bare WTe$_2$. Below 45 K a jump appears in the vicinity of $\pm 1.8$ T, and at lower temperatures a second jump becomes visible at around $\pm 0.5$ T. The positions of the jumps are not affected by $V_g$ (Supplementary SI-3) or by current in the WTe$_2$. Figure 2b compares the conductance (upper panel) with the reflection magnetic circular dichroism (RMCD) signal (lower panel) measured on the same device at 11.8 K (for more temperatures, see Supplementary SI-4). The RMCD signal is roughly proportional to the total out-of-plane magnetization, and the plateaus in it correspond to the four magnetization states of trilayer CrI$_3$ indicated by the schematics beneath. The inset to Fig. 2b is a 2D map of the RMCD signal taken at $B = 0$ during a downwards sweep of $B$. Its uniformity implies that there is a single magnetic domain over most of the WTe$_2$.

It is clear that the conductance jumps occur when the magnetic state of the CrI$_3$ changes. The higher-field jumps accompany transitions between antiferromagnetic (AF) and fully polarized (F) states, which we refer to ↓↑↓ −↓↓↓ and ↑↓↑ −↑↑↑, where the last arrow represents the polarity of the lowest layer, the one in contact with the WTe$_2$. Such spin-flip transitions have little hysteresis. The lower-field jumps accompany transitions between the two antiferromagnetic states, ↓↑↓ −↑↓↑. They show larger hysteresis because the magnetic reconfiguration is more drastic.

The sizes and signs of the conductance jumps are consistent with a simple model in which the WTe$_2$ conduction electrons experience a perpendicular exchange magnetic field, $B_{ex}$, that adds to the external field so that the conductance becomes $G_0(B + B_{ex})$, where $G_0(B)$ is conductance without the CrI$_3$. We assume $B_{ex}$ takes values $\pm B_F$ for ↑↑↑ and ↓↓↓, and $\pm B_{AF}$ for ↑↓↑ and ↓↑↓, respectively. The variation of $B + B_{ex}$ with $B$ is sketched in Fig. 2c. In this model the higher-field jump is between $G_0(B + B_{AF})$ and $G_0(B + B_F)$. Since $G_0(B)$ is roughly linear with similar slope on either side of this jump, as indicated by the dashed lines drawn on the 25 K data in Fig. 2a, we can use the separation of these lines to estimate $B_F - B_{AF} \approx +1$ T. The lower-field jump, which should be between $G_0(B_{AF} + B)$ and $G_0(B - B_{AF})$, cannot be analyzed so simply, but making use of the Onsager symmetry $G_0(B) = G_0(-B)$ we can infer that $B_{AF}$ is much larger than the coercive field, putting it at several Tesla (Supplementary SI-5). If we hypothesize (without theoretical rigor) that the activation energy $E_a$ for the edge is a Zeeman energy associated with $B_{AF}$, i.e., $E_a \approx g\mu_B B_{AF}$, and use a $g$-factor of $g = 4$ estimated from the magnetoresistance (Supplementary SI-6), we obtain $B_{AF} \sim 10$ T, where $\mu_B$ is Bohr magneton. This is similar in magnitude to the exchange field of 13 T found in WSe$_2$/CrI$_3$ heterostructures[10] and of >14 T in graphene/EuS[11].

We next investigate the nonlinear conductance, which yields additional information since it is not constrained by the Onsager symmetry imposed by near-equilibrium conditions. We begin by working at higher temperatures where the linear conductance $G$ is measurable. We apply an ac bias of rms amplitude $V_f$ at frequency $f$ and measure the resulting ac current components at $f$ and $2f$, with $V_f$ chosen such that $I_{2f} << I_{1f}$. If we write $I = GV + \alpha V^2 + \cdots$ then $I_{1f} = GV_f$, and $I_{2f} = \alpha V_f^2/2$ is proportional to the coefficient $\alpha$ which parameterizes the conductance asymmetry between positive and negative bias directions, as indicated in Fig. 3a. Measurements of $I_{1f}$ vs $B$ for device C1, shown in Fig. 3b, match the ac linear conductance measurements in Fig. 2a as expected, exhibiting four jumps. Measurements of $I_{2f}$ at several temperatures, plotted in Fig. 3c, show large jumps corresponding to the ↑↓↑ − ↓↑↓ transitions, detectable up to 40 K, but no discernable features at the ↓↓↓ − ↓↑↓ or ↑↑↑ − ↑↓↑ transitions in which the magnetization of the lowest layer does not flip (Supplementary SI-7 and SI-8). We conclude that the asymmetry parameter $\alpha$ is sensitive to the magnetization, $m$, of the nearest layer of CrI$_3$ but not to that of the next nearest layer.

In these nonlinear measurements the larger bias used could potentially drive some current through the WTe$_2$ bulk, whose activation gap is ~50 meV[14], so we performed an experiment to test whether this bulk current is relevant. Figure 3d shows $I_{2f}$ vs $B$ for a large ac bias ($V_f = 100$ mV) applied to a device C3, which has the contact pattern indicated in the sketches (see Supplementary SI-9 for device and other details). In two-terminal measurements between the outer contact pair (left), $I_{2f}$ shows large jumps. However, when the intervening contact is grounded (right) the jumps almost disappear even though the linear conductance remains substantial. This implies that when current is prevented from flowing along the edge there is no sensitivity to the state of the CrI$_3$. We deduce that in both nonlinear and linear regimes the sensitivity to the magnetic state of the CrI$_3$ is dominated by the sample edges.

Finally, we turn to the nonlinear behavior at lower temperatures, where the linear conductance freezes out. Figure 4 shows two-terminal $I - V$ traces at 1.6 K for bilayer CrI$_3$ device C2 at $B = 0$. The magnetization transitions in bilayer CrI$_3$ follow a different pattern from those in trilayer (Fig. 2b) and occur at different magnetic fields. The bilayer is antiferromagnetic (↑↓ or ↓↑) at low $B$ and flips to a fully polarized state (↑↑ or ↓↓) for $B >\approx 0.9$ T. Here, no current is detected below a threshold bias of ~70 mV in either direction. Above this bias the current depends strongly on $B$ and exhibits hysteresis between two stable states at low fields (upper inset). Correspondingly, either of the two different $I - V$ traces shown (blue and red) can be obtained. RMCD measurements (shown at the bottom of the upper inset) connect them unambiguously with the ↑↓ or ↓↑ states. An RMCD map taken at $B = 0$ (lower inset) shows a uniform magnetic state over most of the WTe$_2$. Note that a finite RMCD signal in the antiferromagnetic state, as seen here, is normal for CrI$_3$ bilayers[29,30] and implies uncompensated magnetization between the two layers, probably in this case related to contact with the WTe$_2$ (for more temperatures, see Supplementary SI-4 and SI-10).

Inspection of Fig. 4 shows that in the ↑↓ state (blue) the current at positive bias is roughly double that at negative bias, i.e., it is strongly nonreciprocal. This difference is hard to understand in terms of an exchange field because reversing the exchange field in going from ↑↓ to ↓↑ does not affect the linear conductance at all at zero field by Onsager symmetry. Importantly, in the ↓↑ state (red) the opposite is true, implying that the dominant part of the nonreciprocal current is connected to the orientation of the magnetization of the lowest CrI$_3$ layer, **m**, relative to the current direction. This is consistent with the corresponding property of $\alpha$ discussed above. A similar nonreciprocal resistance change on magnetization reversal, referred to as unidirectional magnetoresistance (UMR), was reported in magnetic/nonmagnetic 3D TI thin-film heterostructure[5,6]. However, in that case the effect was much smaller and vanished at zero applied field. This is because the spin polarization of the current-carrying states in the 3D TI is in-plane, perpendicular to the magnetization of the Cr dopants at zero field. In order for the magnetization to distinguish opposite in-plane spin polarizations it must be rotated to have an in-plane component using an in-plane applied magnetic field. In our case, the nonreciprocal effect is orders of magnitude larger (of order 100%) and is present at zero applied field. This is allowed because the spin of the 2D TI edge state is not in-plane, and so states with opposite current have an out-of-plane spin polarization component that couples to the magnetization even at zero field.

As a mechanism for this giant nonreciprocal current, we can exclude a current-induced spin-orbit torque effect because no current flows in the insulating CrI$_3$. Another possibility is the anomalous Nernst effect[5], where a temperature gradient $\nabla T$ is created perpendicular to the edge by ohmic heating and induces a voltage along the edge proportional to $\bm{m} \times \nabla T$. Such a transverse temperature gradient seems unlikely to develop because the ohmic heat should be generated within the edge state itself or in the contacts. Another possible mechanism is analogous to the one put forward in Ref 5. Backscattering that opposes current flowing, say, to the right in the helical edge requires spin-flips from "up" to "down" (note that the actual spin alignment axis is presently unknown), and vice versa for current flowing in the opposite direction. The spin flip may be assisted by excitation of magnons within the nearest CrI$_3$ layer. These magnons carry spin opposite to the ferromagnetic polarization of that layer; therefore, one polarization allows more scattering of a right-flowing current, and the opposite allows more scattering of a left-flowing current. In this system the effect can be very large because no small-angle scattering is possible in the 1D helical edge, and also because there is little or no conduction in parallel through the bulk.

In summary, we have observed and studied coupling of the magnetism in insulating layered CrI$_3$ to the edge states of a quantum spin Hall insulator (monolayer WTe$_2$). The results are consistent with the edge states being helical (spin locked to momentum) such that time-reversal symmetry breaking due to the magnetization suppresses their conductance. The effect on the linear edge conductance can be interpreted in terms of an exchange field of order 10 T from the nearest CrI$_3$ layer and a much smaller one, of order $(B_F - B_{AF})/2 = 0.5$ T, from the next-nearest CrI$_3$ layer. The nonlinear conductance shows a large directional asymmetry that depends on the magnetization of only the nearest CrI$_3$ layer, and is thus highly sensitive to the AF ground state.

## References:

1   Chang, C.-Z. *et al.* Experimental observation of the quantum anomalous Hall effect in a magnetic topological insulator. *Science* **340**, 167-170 (2013).
2   Katmis, F. *et al.* A high-temperature ferromagnetic topological insulating phase by proximity coupling. *Nature* **533**, 513-516 (2016).
3   Wei, P. *et al.* Exchange-coupling-induced symmetry breaking in topological insulators. *Phys. Rev. Lett.* **110**, 186807 (2013).
4   Avci, C. O. *et al.* Unidirectional spin Hall magnetoresistance in ferromagnet/normal metal bilayers. *Nat. Phys.* **11**, 570-575 (2015).
5   Yasuda, K. *et al.* Large unidirectional magnetoresistance in a magnetic topological insulator. *Phys. Rev. Lett.* **117**, 127202 (2016).
6   Fan, Y. *et al.* Unidirectional magneto-resistance in modulation-doped magnetic topological insulators. *Nano Lett.* **19**, 692-698 (2019).
7   Lv, Y. *et al.* Unidirectional spin-Hall and Rashba−Edelstein magnetoresistance in topological insulator-ferromagnet layer heterostructures. *Nat. Commun.* **9**, 111 (2018).
8   Deng, Y. *et al.* Magnetic-field-induced quantized anomalous Hall effect in intrinsic magnetic topological insulator MnBi$_2$Te$_4$. arXiv:1904.11468 (2019).
9   Chang, L. *et al.* Quantum phase transition from axion insulator to Chern insulator in MnBi$_2$Te$_4$. arXiv:1905.00715 (2019).
10  Zhong, D. *et al.* Van der Waals engineering of ferromagnetic semiconductor heterostructures for spin and valleytronics. *Sci. Adv.* **3**, e1603113 (2017).
11  Wei, P. *et al.* Strong interfacial exchange field in the graphene/EuS heterostructure. *Nat. Mater.* **15**, 711-716 (2016).
12  Qian, X. *et al.* Quantum spin Hall effect in two-dimensional transition metal dichalcogenides. *Science* **346**, 1344-1347 (2014).
13  Fei, Z. *et al.* Edge conduction in monolayer WTe$_2$. *Nat. Phys.* **13**, 677-682 (2017).
14  Tang, S. *et al.* Quantum spin Hall state in monolayer 1T'-WTe$_2$. *Nat. Phys.* **13**, 683-687 (2017).
15  Wu, S. *et al.* Observation of the quantum spin Hall effect up to 100 kelvin in a monolayer crystal. *Science* **359**, 76-79 (2018).
16  Shi, Y. *et al.* Imaging quantum spin Hall edges in monolayer WTe$_2$. *Sci. Adv.* **5**, eaat8799 (2019).

17    Huang, B. *et al*. Layer-dependent ferromagnetism in a van der Waals crystal down to the monolayer limit. *Nature* **546**, 270-273 (2017).
18    Tokura, Y. *et al*. Nonreciprocal responses from non-centrosymmetric quantum materials. *Nat. Commun.* **9**, 3740 (2018).
19    König, M. *et al*. Quantum spin Hall insulator state in HgTe quantum wells. *Science* **318**, 766-770 (2007).
20    Liu, C.-X. et al. Quantum anomalous Hall effect in $Hg_{1-y}Mn_yTe$ quantum wells. *Phys. Rev. Lett.* **101**, 146802 (2008).
21    Gong, C. & Zhang, X. Two-dimensional magnetic crystals and emergent heterostructure devices. *Science* **363**, eaav4450 (2019).
22    Liu, L. *et al*. Spin-torque switching with the giant spin Hall effect of Tantalum. *Science* **336**, 555-558 (2012).
23    MacNeill, D. *et al*. Control of spin-orbit torques through crystal symmetry in $WTe_2$/ferromagnet bilayers. *Nat. Phys.* **13**, 300-305 (2017).
24    Sajadi, E. *et al*. Gate-induced superconductivity in a monolayer topological insulator. *Science* **362**, 922-925 (2018).
25    Fatemi, V. *et al*. Electrically tunable low-density superconductivity in a monolayer topological insulator. *Science* **362**, 926-929 (2018).
26    Fu, L. & Kane, C. L. Josephson current and noise at a superconductor/quantum-spin-Hall-insulator/superconductor junction. *Phys. Rev. B* **79**, 161408 (2009).
27    Song, T. *et al*. Giant tunneling magnetoresistance in spin-filter van der Waals heterostructures. *Science* **360**, 1214-1218 (2018).
28    Klein, D. R. *et al*. Probing magnetism in 2D van der Waals crystalline insulators via electron tunneling. *Science* **360**, 1218-1222 (2018).
29    Huang, B. *et al*. Electrical control of 2D magnetism in bilayer CrI3. *Nat. Nanotech.* **13**, 544-548 (2018).
30    Jiang, S. *et al*. Controlling magnetism in 2D CrI3 by electrostatic doping. *Nat. Nanotech.* **13**, 549-553 (2018).

**Acknowledgements**: The authors acknowledge Lukasz Fidkowski and Di Xiao for insightful discussions. All the experiments and analysis were supported by NSF DMR grants EAGER 1936697 and MRSEC 1719797. Materials synthesis at UW was partially supported by the Gordon and Betty Moore Foundation's EPiQS Initiative, Grant GBMF6759 to JHC. Materials synthesis at ORNL by MAM was supported by the US Department of Energy, Office of Science, Basic Energy Sciences, Materials Sciences and Engineering Division.

**Author Contributions**: DC and XX conceived the experiment. WZ, ZF, HC, TP and BS fabricated the devices. WZ and ZF performed transport measurements. TS performed magnetic circular dichroism measurements. PM and JC grew the $WTe_2$ crystals. MM grew $CrI_3$ crystals. DC, WZ, XX and ZF wrote the paper with comments from all authors.

**Methods:**

*Device fabrication:* First, graphite and hBN crystals were mechanically exfoliated onto thermally grown SiO$_2$ on a highly doped Si substrate. The thickness of hBN flakes as top and bottom dielectrics are listed in Supplementary Table S1. By using a polymer-based dry transfer technique[31], the few-layer graphite is covered by an hBN flake (bottom hBN). After dissolving the polymer, the hBN/graphite is annealed at 400 ºC for 2 hours. Next, Pt metal contacts (~7 nm) were deposited on the hBN by standard e-beam lithography and metallized in an e-beam evaporator. Then, another step of e-beam lithography and metallization was used to define bond pads (Au/V) connecting to the metal contacts and the graphite gate. CrI$_3$ and WTe$_2$ crystals were exfoliated in a glove box (O$_2$ and H$_2$O concentrations < 0.5 ppm). CrI$_3$ flakes from bilayer to four-layer and monolayer WTe$_2$ flakes were optically identified. A CrI$_3$ flake was picked up under another hBN flake (top hBN), followed by a pick-up of the monolayer WTe$_2$ flake. The stack was then put down on the Pt contacts in the glove box. Finally, the polymer was quickly dissolved in chloroform (~1 minute).

*Electrical measurements:* Electrical measurements were carried out in an Oxford He-4 VTI cryostat with temperature down to 1.6 K and magnetic field up to 14 T. A 1 mV a.c. excitation at 101 Hz was applied for linear responses. For second harmonic responses, a 15-100 mV a.c excitation at 101 Hz was applied, while at the same time we also connected a 30 $\mu$F capacitor in series with the device.

*Reflective magnetic circular dichroism measurements:* Reflective magnetic circular dichroism measurements were performed in a closed-cycle cryostat (attoDRY 2100) with a base temperature of 1.6 K and an out-of-plane magnetic field up to 9 T. A 632.8 nm helium–neon laser was used to probe the device at normal incidence with a fixed power of 100 nW. The standard lock-in measurement technique used to measure the RMCD signal closely followed the previous magneto-optical Kerr effect and RMCD measurements of the magnetic order in atomically thin CrI$_3$[17,27].

*References:*

31   Zomer, P. J., Guimaraes, M. H. D., Brant, J. C., Tombros, N. & van Wees, B. J. Fast pick up technique for high quality heterostructures of bilayer graphene and hexagonal boron nitride. *Appl. Phys. Lett.* **105**, 013101 (2014).

**Main figures:**

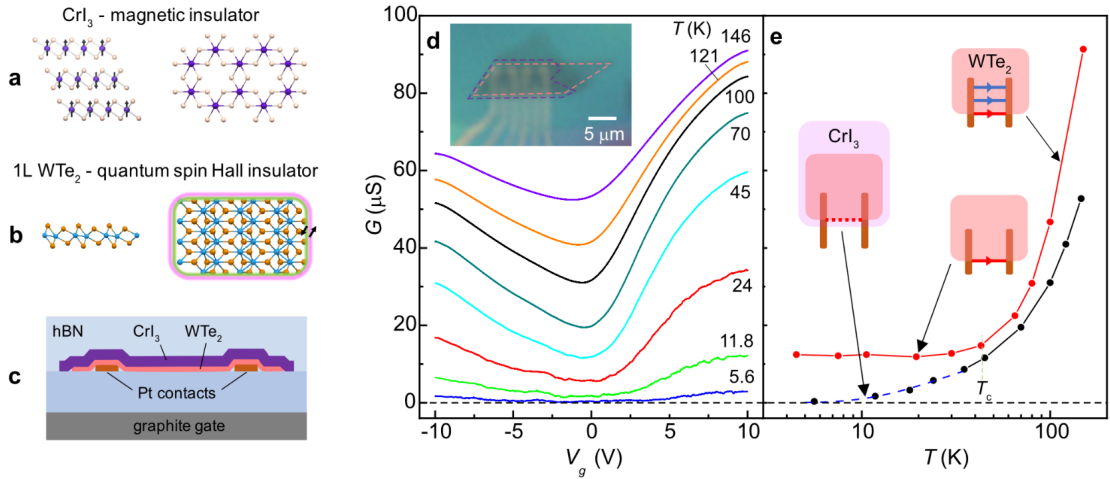

**Figure 1 | Characteristics of a CrI$_3$/WTe$_2$ device with no applied field. a,** Side view indicating the layered antiferromagnetic order in CrI$_3$ below $T_c \approx 45$ K, and top view of monolayer structure (Cr: purple, I: yellow). **b,** Side and top views of the structure of monolayer (1L) WTe$_2$ (W: blue, Te: orange). The presence of helical edge states is indicated in cartoon form. **c,** Schematic device cross-section. **d,** Gate dependence of the linear conductance between two adjacent contacts (separated by 0.8 μm) at different temperatures for device C1, which has trilayer CrI$_3$. The capacitively induced doping per gate voltage is $1.2 \times 10^{12}$ cm$^{-2}$/V. Inset: optical image of device C1; the trilayer CrI$_3$ and 1L WTe$_2$ flakes are outlined by purple and pink dashed lines, respectively. **e,** Temperature dependence of the minimum conductance for device C1 (black) and for bare (no CrI$_3$) 1L WTe$_2$ device M1 (red). The insets indicate the relevance of bulk and edge currents in each case.

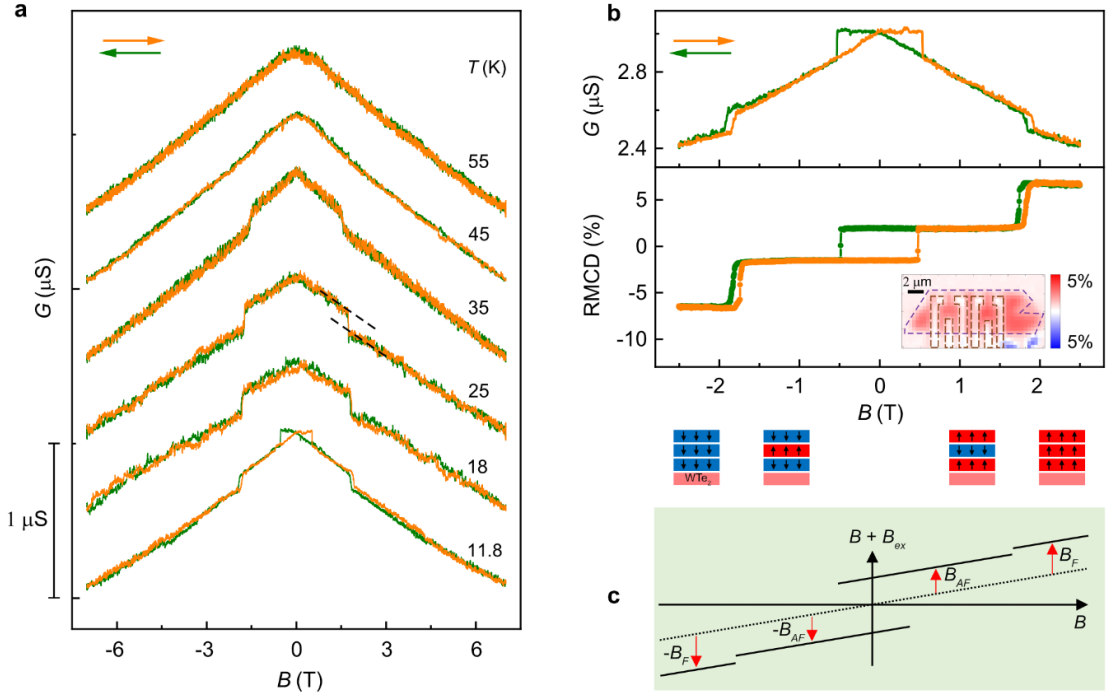

**Figure 2 | Conductance jumps and magnetic state changes in an applied magnetic field. a**, Linear conductance (measured using 1 mV ac bias) vs. out-of-plane magnetic field $B$ at $V_g = -0.5$ V for device C1 at the indicated temperatures. The dashed lines indicate the effective shift of the characteristic that occurs at one of the jumps (see text). The traces are vertically offset for clarity; the conductance at $B = 0$ can be read off in Fig. 1e. **b**, Upper: conductance, and lower: RMCD (reflection magnetic circular dichroism) signal as a function of $B$ (10 mV ac bias, $V_g = -1$ V, $T = 11.8$ K). Inset: spatial map of RMCD signal at $B = 0$ after reducing $B$ from +2.5 T. The boundaries of the Pt contacts and trilayer CrI$_3$ are indicated by purple and dark yellow dashed lines, respectively. Beneath are schematics of the corresponding magnetic states of the trilayer CrI$_3$ (blue and red for down and up polarizations) atop 1L WTe$_2$ (pink). **c**, Schematic variation of the sum of the real and exchange magnetic fields used to interpret the behavior of $G$ in panel b.

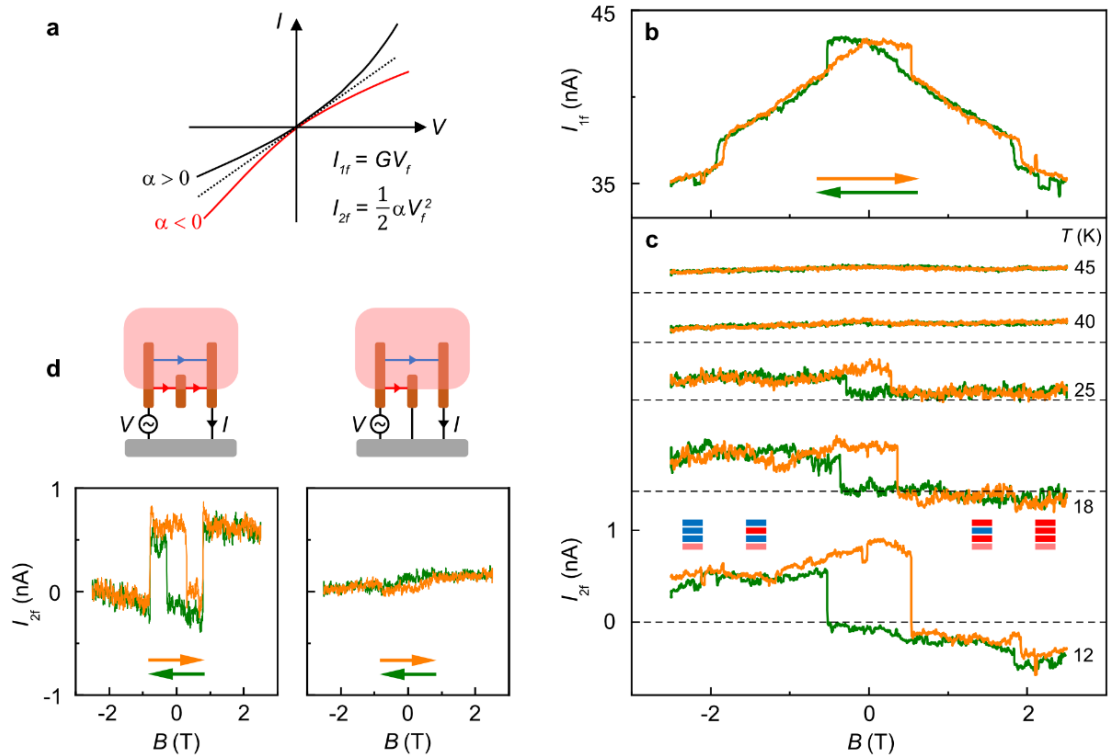

**Figure 3 | Nonlinear current measurements. a,** Sketch indicating the relationship of the first- and second-harmonic current components to the asymmetry in the $I - V$ which is parameterized by quadratic coefficient $\alpha$. **b**, $I_{1f}$ vs $B$ for device C1 at $V_g = -0.5$ V and $T = 12$ K with ac voltage bias $V_f = 15$ mV. **c**, $I_{2f}$ vs $B$ measured under the same conditions (lowest traces) and at several other temperatures as labelled. The traces are vertically offset for clarity and the dashed horizontal lines show the zero level for each temperature. The schematics are repeated from Fig. 2c. **d**, Second harmonic current vs $B$ for device C3 at $V_g = 0.1$ V, $T = 27$ K and $V_f = 100$ mV, compared between the two different measurement configurations as indicated. Bulk and edge currents are signified by blue and red arrows respectively.

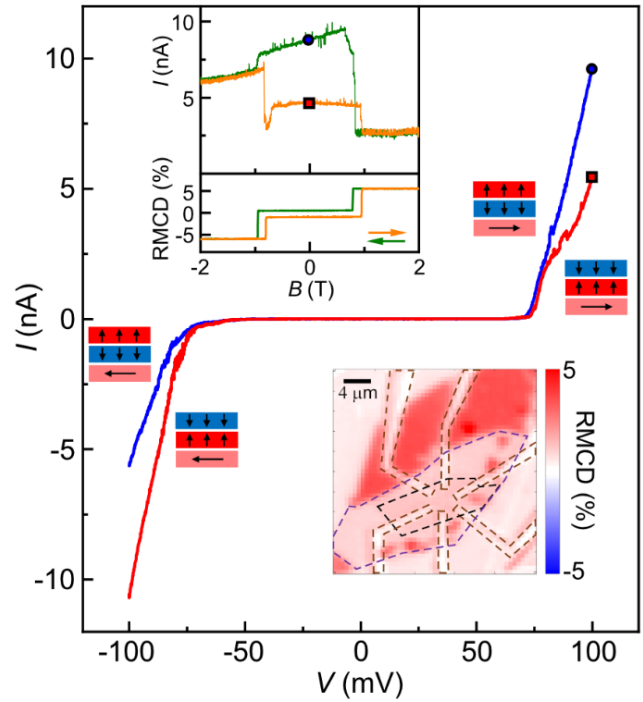

**Figure 4 | Large nonreciprocal current controlled by the antiferromagnetic state.** *I-V* traces at zero magnetic field at 1.6 K and $V_g = 0$ V for device C2, which has bilayer $CrI_3$. The schematics indicate the magnetization state of the $CrI_3$ and the current flow direction in the 1L $WTe_2$. Upper inset: current at 100 mV dc bias (corresponding to the points marked with symbols in the *I-V* traces) and RMCD (reflection magnetic circular dichroism) signal vs magnetic field. Lower inset: spatial map of RMCD signal at 0 T after reducing from +2.5 T. The current is measured between the upper two contacts. The boundaries of bilayer $CrI_3$, Pt contacts, and $WTe_2$ are outlined with purple, dark yellow and black dashed lines, respectively.

# Magnetic proximity and nonreciprocal current switching in a monolayer WTe$_2$ helical edge

Wenjin Zhao[1], Zaiyao Fei[1], Tiancheng Song[1], Han Kyou Choi[1], Tauno Palomaki[1], Bosong Sun[1], Paul Malinowski[1], Michael A. McGuire[3], Jiun-Haw Chu[1], Xiaodong Xu[1,2,*], David H. Cobden[1,*]

[1]Department of Physics, University of Washington, Seattle, WA, USA
[2]Department of Materials Science and Engineering, University of Washington, Seattle, WA, USA
[3]Materials Science and Technology Division, Oak Ridge National Laboratory, Oak Ridge, TN, USA
*Corresponding author: xuxd@uw.edu, cobden@uw.edu

## SI-1. Preparation and characterization of WTe$_2$/CrI$_3$ devices

We measured devices with three different configurations: (1) monolayer WTe$_2$ covered by few-layer CrI$_3$ (C1, C2, C3); (2) monolayer WTe$_2$ partially covered by bilayer CrI$_3$ (C4); (3) double gated monolayer WTe$_2$ (M1, M2). The fabrication of the first type of device is described in the Methods section. The second and third types are similar. Figure S1c shows the optical image of a typical monolayer WTe$_2$/few-layer CrI$_3$ device (C2).

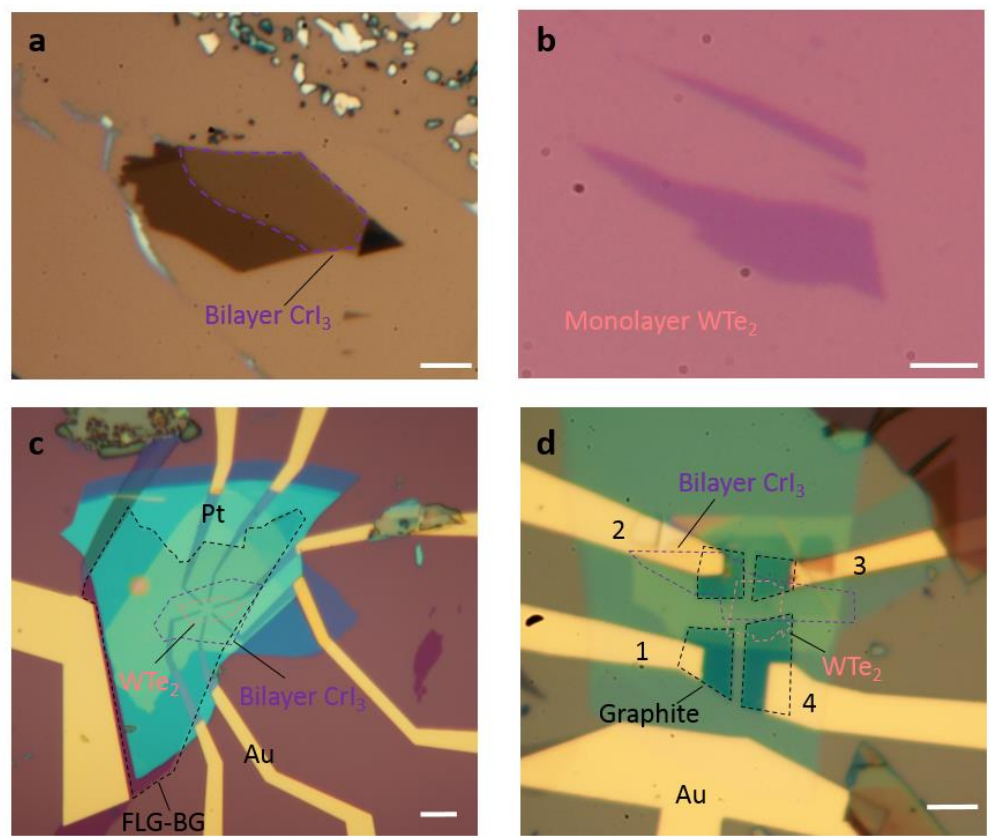

**Figure S1 | Monolayer WTe$_2$/few-layer CrI$_3$ devices. a,** Optical microscope image of bilayer CrI$_3$. Scale bar: 3 μm. **b,** Optical microscope image of monolayer WTe$_2$. Scale bar: 2 μm. **c,** Optical

microscope image of device C2. Scale bar: 10 μm. **d**, Optical microscope image of device C4. Etched graphite was used for contacts in C4, and the monolayer $WTe_2$ is only partially covered by bilayer $CrI_3$. For example, between contacts 1 and 4 the $WTe_2$ edge is not covered by bilayer $CrI_3$. The pink and purple dashed lines outline the monolayer $WTe_2$ and the bilayer $CrI_3$, respectively. Scale bar: 10 μm.

Table S1 lists the thicknesses of the top and bottom hBN and corresponding areal geometric capacitance $C_g$. The change in electron-hole density imbalance is approximately $n_e = C_g V_g/e$, where $C_g = \epsilon_{hBN}\epsilon_0/d_b$, $\epsilon_{hBN} \approx 4$ is the dielectric constant of hBN[1], and $d_b$ is the thicknesses of the bottom hBN flakes.

| Device label | $WTe_2$ | $CrI_3$ | top hBN (nm) | bottom hBN (nm) | $C_g$ $(1 \times 10^{-3} \text{ F/m}^2)$ |
|---|---|---|---|---|---|
| C1 | Monolayer | Trilayer | 20.9 | 18.3 | 1.9 |
| C2 | Monolayer | Bilayer | 16.0 | 37.0 | 1.0 |
| C3 | Monolayer | Four-layer | 17.2 | 7.0 | 5.1 |
| C4 | Monolayer | Bilayer | 12.0 | 23.5 | 1.5 |
| M1 | Monolayer | NA | 11.4 | 14.0 | 2.5 |
| M2 | Monolayer | NA | 9.2 | 17.5 | 2.0 |

Table S1 | Thickness of the $WTe_2$, $CrI_3$, and top and bottom hBN in each device. The hBN thicknesses were obtained from AFM images. The thicknesses of $WTe_2$ and $CrI_3$ were determined from their optical contrast.

## SI-2. Characteristics of an additional device (C2, bilayer $CrI_3$) with no applied field and comparison of *I-V* for edges with and without $CrI_3$ in device C4

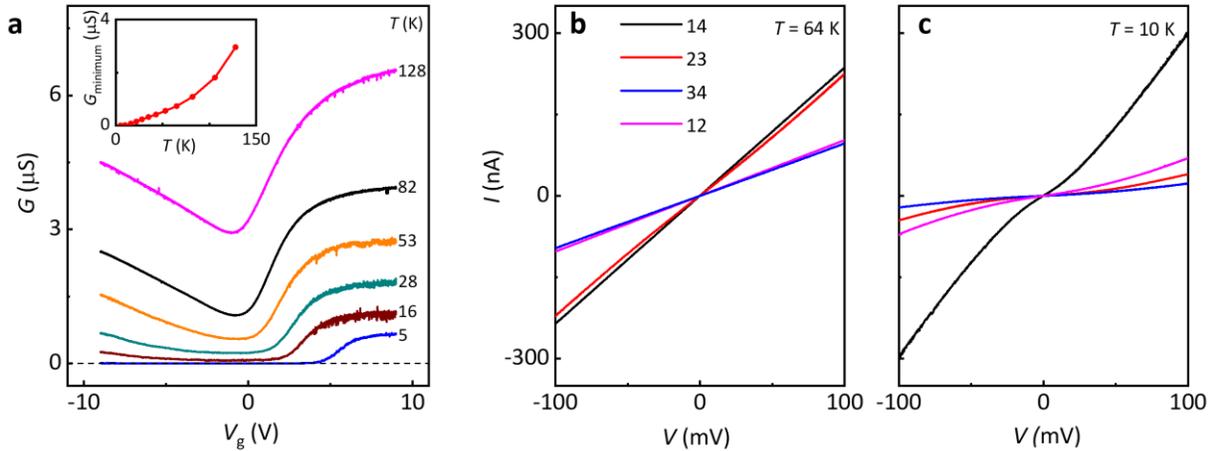

**Figure S2 | a,** Gate dependence of the linear conductance between two adjacent contacts at different temperature for device C2. The temperature dependence of the minimum conductance is drawn in the inset. **b** and **c,** Comparison of *I-V* characteristics for channels with and without $CrI_3$. **b**, at 64 K. **c**, at 10 K.

In device C4 the WTe$_2$ is partially covered by bilayer CrI$_3$ as shown in Fig. S1d. Edge 14 is not covered at all by bilayer CrI$_3$, edges 23 and 34 are totally covered, and edge 12 is partially covered. The lengths of edges 14 and 23 are similar while edges 12 and 34 are longer. Above the $T_c$ of bilayer CrI$_3$, which is ~45 K, the *I-V* curves for pairs 14 and 23 are similar and the *I-V* curves for pairs 34 and 12 are similar, as shown in Fig. S2b. Below $T_c$, the current for the uncovered edge, pair 14, is much higher than for the other three edges which are all either partially or fully covered by bilayer CrI$_3$, as shown in Fig. S2c.

It is possible for the current to go all around the outside of the sample under the uncovered edge. For this reason, we always ground at least one other contact to eliminate this possibility.

## SI-3. Different gate voltages in device C1

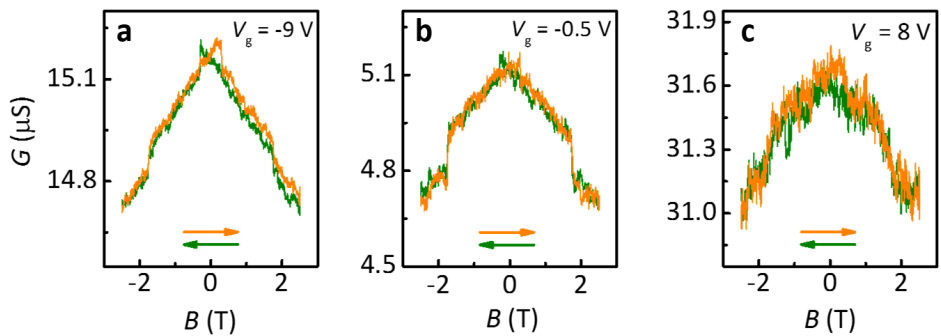

**Figure S3 | Comparison of magnetoconductance at different gate voltages. a-c,** Conductance of monolayer WTe$_2$ in device C1 at $T = 11.8$ K as a function of an out-of-plane magnetic field for $V_g$ = -9, -0.5 and +8 V, respectively. When the bulk of the WTe$_2$ is conducting, at $V_g$ = -9 and +8 V, conductance jumps can still be seen at the same magnetic field but they are not as clear as in the edge-dominated regime at $V_g = -0.5$ V.

# SI-4. Temperature dependence of RMCD signal

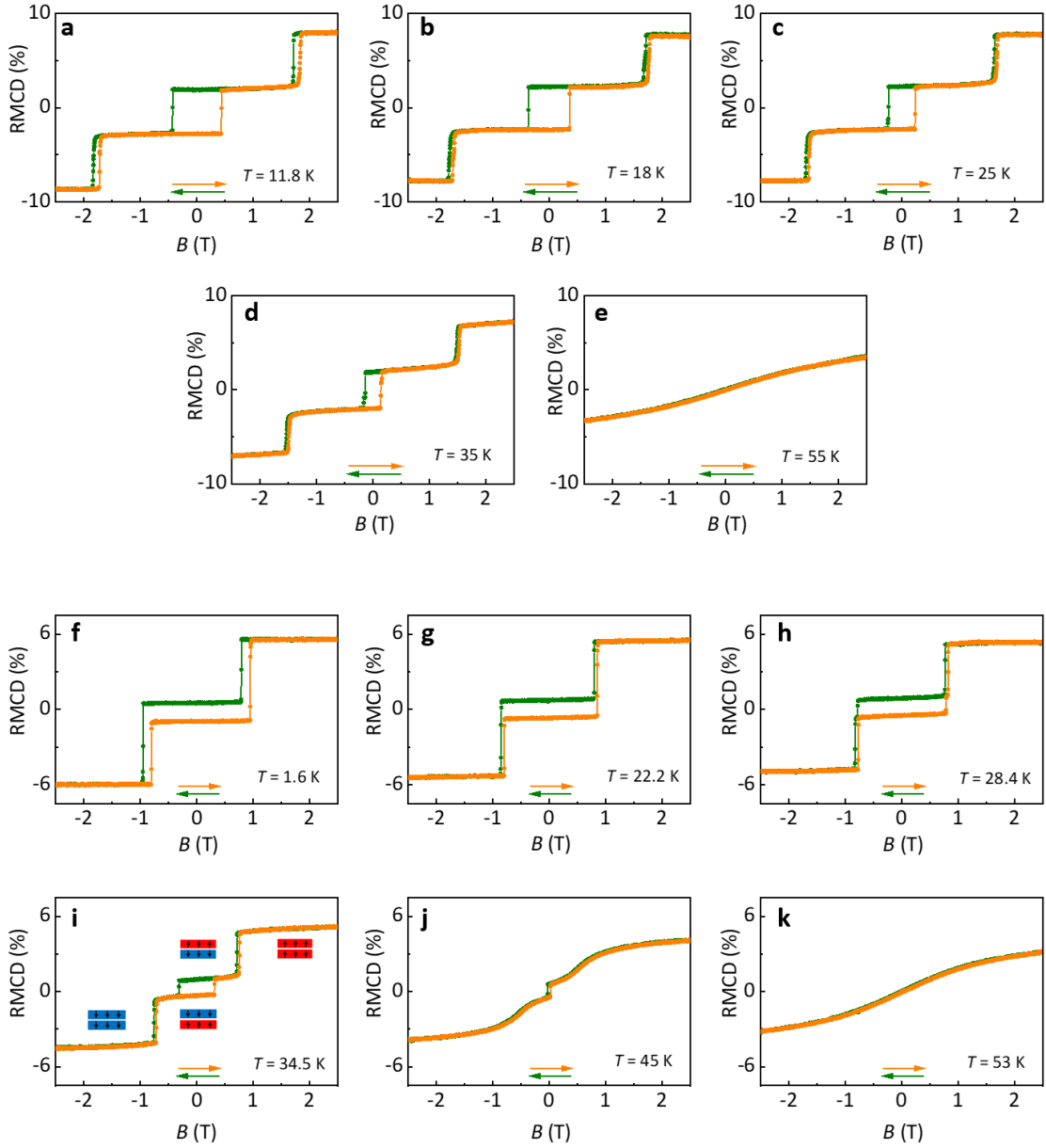

**Figure S4 | RMCD signal as a function of an out-of-plane magnetic field at different temperatures as indicated. a-e,** for device C1 (trilayer CrI$_3$). **f-k,** for device C2 (bilayer CrI$_3$).

## SI-5. Estimate of the coercive field for the transition ↓↑↓ − ↑↓↑ in trilayer CrI$_3$ device C1

Figure S5a shows sketches of the form of the hysteresis loop expected for the cases $B_{AF} \ll B_c$ (top) and $B_{AF} \gg B_c$ (bottom), where $B_c$ is the coercive field. The blue lines schematically show $G_0(B+B_{AF})$ and $G_0(B-B_{AF})$ for ↑↓↑ and ↓↑↓, respectively. $G$ jumps between the two branches at $B_c$ when the transition occurs. The shape of the hysteresis loop varies somewhat in shape with gate voltage. However, as for the example in Fig. S5b, it is always more compatible with the case $B_{AF} \gg B_c$, i.e., $B_{AF} \gg 0.5$ T.

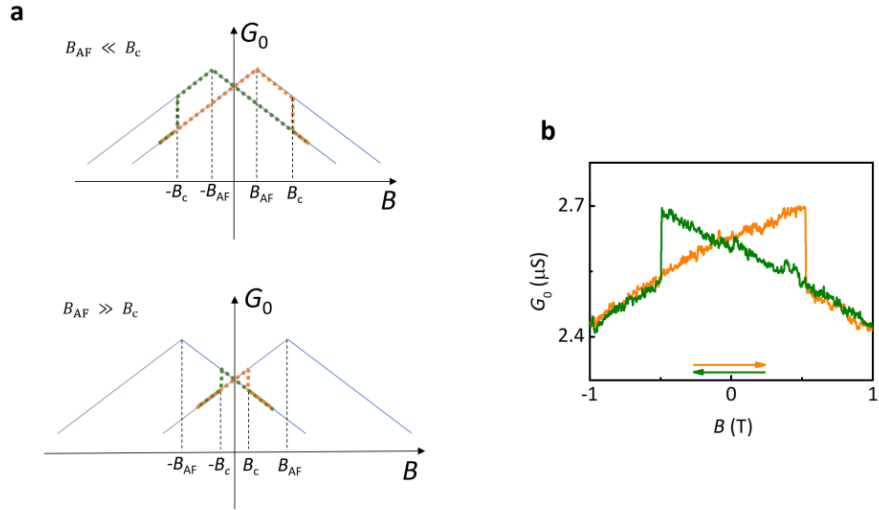

**Figure S5 | Comparison of $B_{AF}$ and $B_c$ a,** Sketches shown the shape of hysteresis loop for $B_{AF} \ll B_c$ (top) and $B_{AF} \gg B_c$ (bottom). **b,** Hysteresis loop (measured using 10 mV ac bias) at $V_g$ = -0.5 V for device C1 at 11.8 K.

## SI-6. Estimating the g-factor of the edge state electrons in out-of-plane magnetic field

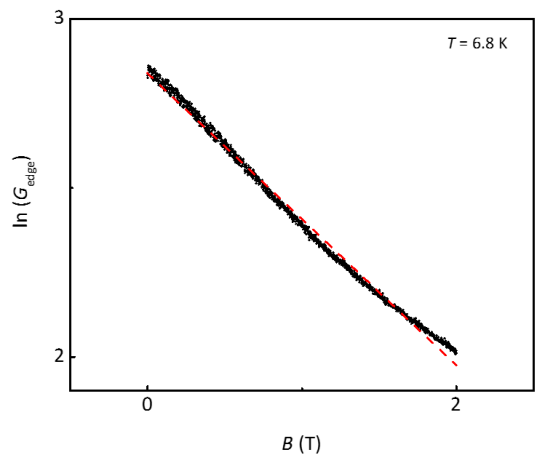

**Figure S6 | ln($G_{edge}$) as function of an out-of-plane magnetic field - device M2.** Using $G_{edge} = G_0 e^{-\frac{g\mu_B B}{k_B T}}$ to fit data (black dots). Red dashed line is the best fit with $g \sim 4$.

## SI-7. Zoom-in of first and second harmonic responses at 40 K and nonreciprocal resistance by measuring differential conductance with a dc bias

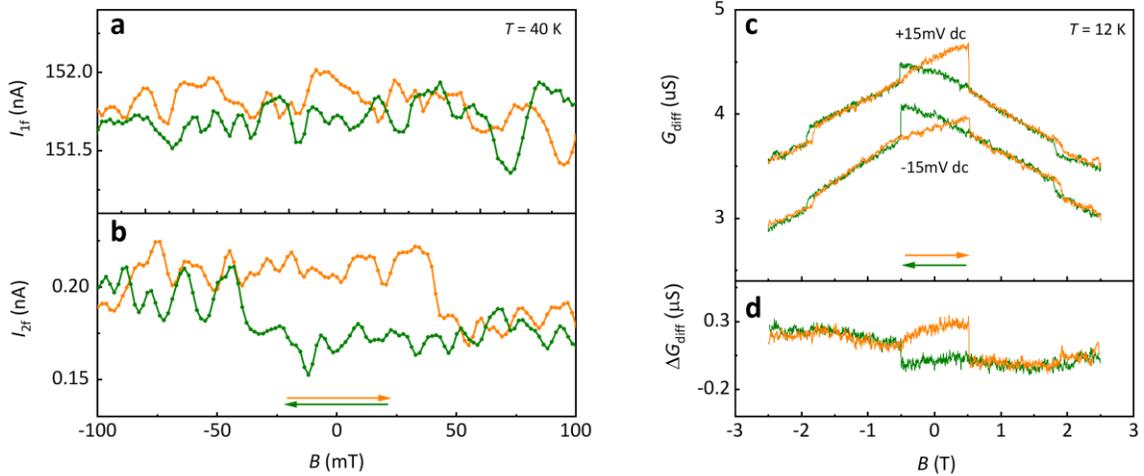

**Figure S7 | a-b,** Zoom-in of 1f (**a**) and 2f (**b**) responses at 40 K for device C1 at low magnetic field. A clear hysteresis loop only shows in the 2f response at 40 K. **c,** Comparison of differential conductance for opposite signs of the dc bias at 12 K. Differential conductance ($G_{diff}$) as a function of an out-of-plane magnetic field at +15 mV and -15 mV dc bias respectively. The -15 mV traces are shifted vertically for clarity. **d,** Difference between differential conductance ($\Delta G_{diff}$) at +15 mV and -15 mV bias as a function of an out-of-plane magnetic field. $G_{diff}$ shows four jumps while $\Delta G_{diff}$ shows only two jumps corresponding to the transitions between ↑↓↑ and ↓↑↓. As discussed in the main text, this implies that $\Delta G_{diff}$ is mainly controlled by the magnetization of the nearest layer of $CrI_3$.

## SI-8. First and second harmonic responses at different gate voltages

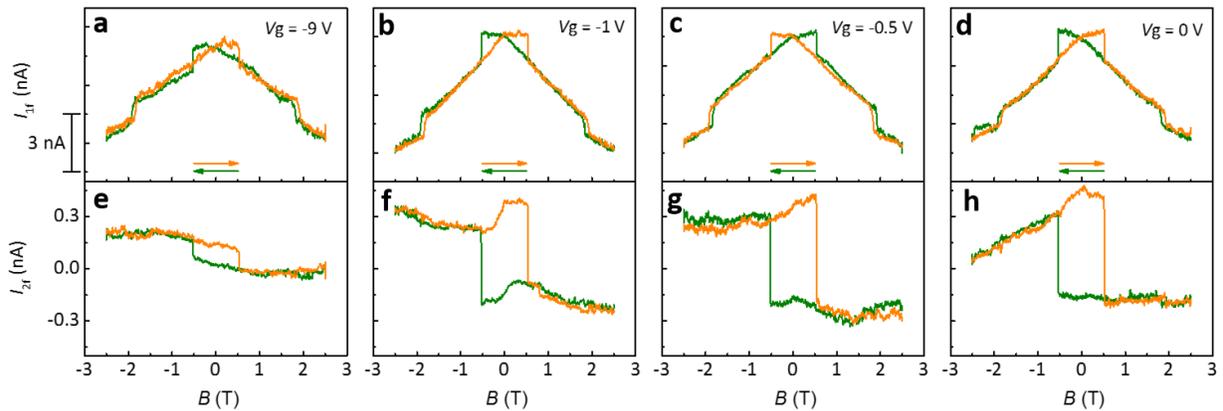

**Figure S8 | Gate dependence of 1f and 2f response with 10mV ac bias at T = 11.8 K in device C1. a-d,** First harmonic response as a function of out-of-plane magnetic field at $V_g$ = -9, -1, -0.5, 0 V, respectively. **e-h,** Second harmonic response as a function of out-of-plane magnetic field at $V_g$ = -9, -1, -0.5, 0 V, respectively.

## SI-9. Separation of bulk and edge of monolayer WTe$_2$

We performed two experiments that indicate that even in the nonlinear regime the sensitivity to the magnetic state of the CrI$_3$ is only at the sample edges. First, Fig. S9a shows $I_{1f}$ (left panel) and $I_{2f}$ (right panel) vs $B$ for a large ac bias of $V_f = 100$ mV applied to device C3. This device had nominally 4-layer CrI$_3$ that behaved like 2-layer CrI$_3$, suggesting that two of the layers were damaged/oxidized. It has a contact pattern indicated in the sketches on the left. In two-terminal measurements between the outer contacts (top row) both $I_f$ and $I_{2f}$ show large jumps. However, when the intervening contact is grounded (bottom row) the jumps almost disappear, $I_f$ becomes almost independent of $B$ though it remains substantial, and $I_{2f}$ almost vanishes. This implies that when current is prevented from flowing along the edge, some still flows through the bulk but it has little or no sensitivity to the magnetic state of the CrI$_3$. Second, Fig. S9b compares measurements of the current resulting from a large dc bias of 100 mV in device C2 for different contact configurations. For adjacent contacts (left) there is large asymmetry and hysteresis around $B = 0$, whereas for contacts on opposite edges (right) the asymmetry and hysteresis are almost absent. This contrast is inexplicable by any bulk current flow phenomenon we can postulate, but can be explained if there is approximate cancellation of nonlinear edge effects on opposite edges of the sample.

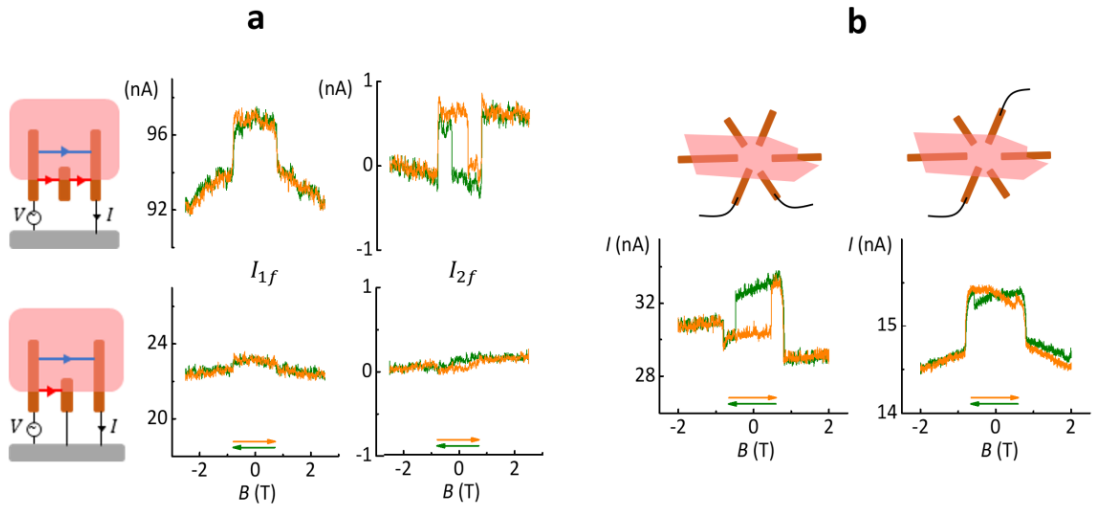

**Figure S9 | Evidence that the magnetic sensitivity is associated only with edge current. a,** Comparison of first and second harmonic current response vs magnetic field for device C3 in the two measurement configurations indicated on the left ($V_f = 100$ mV ac bias and $T = 27$ K). **b,** dc current vs magnetic field for device C2 at 100 mV dc bias and $T = 28$ K for the measurement configurations indicated above.

## SI-10. Temperature dependence of inner jumps in device C2

Between 25 K and 45 K inner jumps (as labeled in Fig. S10a) show up in device C2, consistent with the RMCD measurements in Fig. S4f-k. This implies an uncompensated magnetization between the two layers, possibly related to contact with the $WTe_2$ which may for example induce charge transfer.

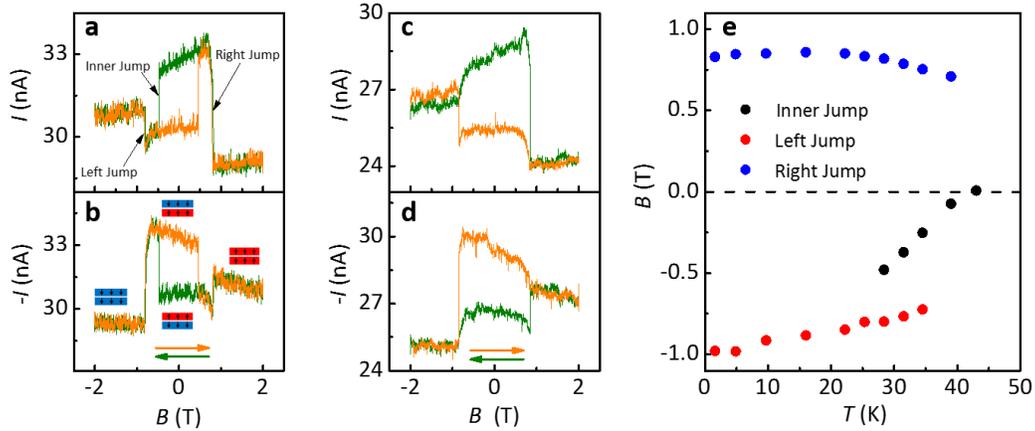

**Figure S10 | Comparison of current at different signs of bias and coercive field as a function of *T*. a, b,** Current as a function of an out-of-plane magnetic field at $T = 28.4$ K for +100 mV dc bias and -100 mV dc bias, respectively. **c, d,** Current as a function of an out-of-plane magnetic field at $T = 22.2$ K for +100 mV dc bias and -100 mV dc bias, respectively. **e**, Coercive field of different jumps labeled in **a** as a function of *T*.

## References:

1    Dean, C. R. *et al.* Boron nitride substrates for high-quality graphene electronics. *Nat. Nanotech.* **5**, 722-726 (2010).